**Nonlinear Processes in Geophysics**

# Surface mixing and biological activity in the four Eastern Boundary Upwelling Systems

V. Rossi[1,2], C. López[2], E. Hernández-García[2], J. Sudre[1], V. Garçon[1], and Y. Morel[3]

[1]Laboratoire d'Études en Géophysique et Océanographie Spatiale, CNRS, Observatoire Midi-Pyrénées, 14 avenue Edouard Belin, Toulouse, 31401 Cedex 9, France
[2]Instituto de Física Interdisciplinar y Sistemas Complejos IFISC (CSIC-UIB), Campus Universitat de les Illes Balears, 07122 Palma de Mallorca, Spain
[3]Service Hydrographique et Océanographique de la Marine, (SHOM), 42 avenue Gaspard Coriolis, 31057 Toulouse, France

**Abstract.** Eastern Boundary Upwelling Systems (EBUS) are characterized by a high productivity of plankton associated with large commercial fisheries, thus playing key biological and socio-economical roles. Since they are populated by several physical oceanic structures such as filaments and eddies, which interact with the biological processes, it is a major challenge to study this sub- and mesoscale activity in connection with the chlorophyll distribution. The aim of this work is to make a comparative study of these four upwelling systems focussing on their surface stirring, using the Finite Size Lyapunov Exponents (FSLEs), and their biological activity, based on satellite data. First, the spatial distribution of horizontal mixing is analysed from time averages and from probability density functions of FSLEs, which allow us to divide each areas in two different subsystems. Then we studied the temporal variability of surface stirring focussing on the annual and seasonal cycle. We also proposed a ranking of the four EBUS based on the averaged mixing intensity. When investigating the links with chlorophyll concentration, the previous subsystems reveal distinct biological signatures. There is a global negative correlation between surface horizontal mixing and chlorophyll standing stocks over the four areas. To try to better understand this inverse relationship, we consider the vertical dimension by looking at the Ekman-transport and vertical velocities. We suggest the possibility of a changing response of the phytoplankton to sub/mesoscale turbulence, from a negative effect in the very productive coastal areas to a positive one in the open ocean. This study provides new insights for the understanding of the variable biological productivity in the ocean, which results from both dynamics of the marine ecosystem and of the 3-D turbulent medium.

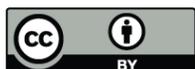

*Correspondence to:* V. Rossi
(vincent.rossi@legos.obs-mip.fr)

## 1 Introduction

Although they represent only a very small fraction of the total surface of the world's ocean, the Eastern Boundary Upwelling Systems (EBUS) are the most productive regions of the world due to their important coastal biological productivities which support large commercial fisheries, up to 20% of the global fish catch (Pauly and Christensen, 1995). They include the Canary (CUS) and the Benguela upwelling systems (BUS) in the Atlantic Ocean and the Peru/Chile (or Humboldt HUS) and California upwelling systems (CalUS) in the Pacific Ocean. Under the action of wind from quasi-stationary high pressure cells over the subtropical ocean basins, a surface uprising of deep cold water rich in nutrients occurs over continental shelves almost all year long. This process explains the high primary production in these regions which constitutes the base of a highly dynamical and rich food chain. Roughly, the intensity of coastal upwelling is modulated by the force and direction of the wind, by the local topography and by the ambient oceanic characteristics.

These EBUS are spatially and temporally heterogeneous from both a physical and biological point of view. The development of diverse structures such as intense fronts, coastal plumes in retention areas, offshore filaments and eddies interplays with the complex spatial distribution of phytoplankton. This mesoscale and sub-mesoscale oceanic turbulence is known to strongly modulate the structure, biomasses and rates of marine ecosystems, since it can stimulate the primary productivity (McGillicuddy et al., 1998; Oschlies and Garçon, 1998), affect plankton community composition (Owen, 1981; Kang et al., 2004; Mackas et al., 2005) and play a significant role in exchange processes in the transitional area between the productive coastal zone and the oligotrophic open ocean by transporting organic matter and marine organisms from the coast to the open ocean (Moore et al., 2007). This latter mechanism, i.e. the large coastal productivity and its export to the inner ocean via filament formation, identifies them as key regions in the global marine element cycles, such as carbon and nitrogen (Mackas et al., 2006).

While sharing common bio-physical characteristics, their biological productivity is highly variable and governed by diverse factors and their interaction, which are still poorly understood. Several previous comparative studies investigated these major environmental factors and leading physical processes that may control it. When considering all EBUS together, Carr and Kearns (2003) showed that phytoplankton productivity results from a combined effect of large-scale circulation and local factors. Patti et al. (2008) suggested that several driving factors, as nutrients concentration, light availability, shelf extension and among others a surface turbulence proxy from a wind-mixing index, must be taken into account when investigating the phytoplankton biomass distribution. Globally, their statistical study pointed out that all these factors are playing a role whereas they are acting at different levels on the productivity. It is then highly relevant to consider an original Lagrangian measure of mixing for comparative approach among EBUS.

The aim of this study is first to quantify and compare the mixing activity in the EBUS using the technique of the Finite-Size Lyapunov Exponents. The spatial distribution and the temporal evolution of the mixing and stirring activity is analysed. The link between turbulence and chlorophyll concentration (as a proxy for biological activity) is then investigated, leading to propose some underlying mechanisms behind the relationship revealed. Finally, we discuss previous comparative approaches performed among these EBUS with new insights from the present mixing analysis.

## 2 Methods

The basic ingredients of our comparative analysis are satellite data of the marine surface including a two dimensional velocity field and chlorophyll concentration data as a proxy for biological activity and a specific numerical tool employed to analyze these data. We quantify horizontal transport processes by the Lagrangian technique of the Finite Size Lyapunov Exponents (FSLEs) (Aurell et al., 1997), which is specially suited to study the stretching and contraction properties of transport in geophysical data (d'Ovidio et al., 2004). The calculation of the FSLE goes through computing the time, $\tau$, at which two fluid particles initially separated by a distance $\delta_0$ reach a final separation distance $\delta_f$, following their trajectories in a 2 D velocity field. At position $x$ and time $t$ the FSLE is given by: $\lambda(x, t, \delta_0, \delta_f) = \frac{1}{\tau} \log \frac{\delta_f}{\delta_0}$. We are in fact considering the four neighbors of each gridpoint and we selected the orientation of maximum separation rate (fastest neighbor to reach the final separation distance). The equations of motion that describe the horizontal evolution of particle trajectories are computed in longitudinal and latitudinal spherical coordinates ($\phi, \theta$, measured in degrees; $\delta_0$ and $\delta_f$ are also measured in degrees): $\frac{d\phi}{dt} = \frac{u(\phi, \theta, t)}{R \cos \theta}$, $\frac{d\theta}{dt} = \frac{v(\phi, \theta, t)}{R}$. $u$ and $v$ represent the eastward and northward components of the surface velocity field, and $R$ is the radius of the Earth. Numerical integration is performed by using a standard fourth-order Runge-Kutta scheme with an integration time step of $dt = 1$ day. Spatiotemporal interpolation of the velocity data is achieved by bilinear interpolation. We follow the trajectories for 300 days, so that if $\tau$ gets larger than 300 days, we define $\lambda = 0$. FSLEs depend critically on the choice of two length scales: the initial separation $\delta_0$ and the final one $\delta_f$. d'Ovidio et al. (2004) argued that $\delta_0$ has to be close to the intergrid spacing among the points $x$ on which FSLEs will be computed, which is $\delta_0 = 0.025°$. On the other hand, since we are interested in mesoscale structures, $\delta_f$ is chosen equal to 1°, implying a separation distance of about 110 km close to the equator. In this respect, the FSLEs represent the inverse time scale for mixing up fluid parcels between the small-scale grid and the characteristic scales of eddies in these upwelling areas. Choosing slightly different values for $\delta_f$ does not alter qualitatively our results, the main pattern and averages remain the same. To sum up, maps of FSLE are computed monthly for the period June 2000 to June 2005 on all points of a latitude-longitude grid with a spacing of $\delta_0 = 0.025°$. An alternative tool to FSLE is the Finite-Time Lyapunov exponents (Haller, 2001; Beron-Vera et al., 2008) but we expect that similar results would be obtained by this last technique for the present spatial and temporal scales. This is so because we use a value of $\delta_0$ smaller than the typical structures in the velocity field, so that FSLE is close to the value of the local Lyapunov exponent and thus of the FTLE at large times (Aurell et al., 1997; Artale et al., 1997). The time integration of the particle trajectories can be performed in two different ways: forward or backward in time. In a typical snapshot of the backwards-in-time dynamics, the maximum values of FSLE organize in lines which are a good approximation for the areas of maximal convergence. On the other hand, FSLE calculated with the forward-in-time integration exhibit large values in the regions of maximal divergence. The line-shaped regions of maximal convergence (divergence) approximate the so-called unstable (stable) manifolds of the hyperbolic trajectories in the flow (Boffeta et al., 2001; Koh and Legras, 2002; d'Ovidio et al., 2004). As a consequence, these ridges, i.e. lines of maximum separation or convergence rates, move with the flow as if they were material lines and thus delineate fluid domains with quite distinct origin and characteristics. Although it would be good to have for ridges in FSLEs some rigorous analysis, of the type of Shadden et al. (2005) for ridges in FTLEs, putting it in a firmer mathematical basis and identifying its limits of validity, there is ample numerical and theoretical evidence confirming this behavior (Koh and Legras, 2002; Lehahn et al., 2007; d'Ovidio et al., 2009). We focus in this work on the backward-in-time dynamics since FSLEs' lines have a clear interpretation as fronts of passive scalars driven by the flow (d'Ovidio et al., 2009). These lines strongly modulate the fluid motion since when reaching maximum values, they act as transport barriers for particle trajectories thus constituting a powerful tool for predicting fronts generated by passive advection, eddy boundaries, material filaments, etc. In a different set of papers (d'Ovidio et al., 2004, 2009; Lehahn et al., 2007; Rossi et al., 2008), the adequacy of FSLE to characterize horizontal mixing and transport structures in the marine surface has been demonstrated as well as its usefulness when correlating with tracer fields like temperature or chlorophyll. Related Lagrangian diagnostics (FTLEs) have even been used to understand harmful algae development (Olascoaga et al., 2008). In addition, spatial averages of FSLEs can define a measure of horizontal mixing in a given spatial area, the larger this spatial average the larger the mixing activity. Following these studies, we will use in this work the FSLE as an analysis tool of the horizontal mixing activity of the surface ocean and will highlight similarities and differences both at a hydrodynamic and biological level.

We study the transitional area of exchange processes between the shelf and offshore in the open ocean. The filaments in the fluctuating boundary between the upwelling and the edge of the oligotrophic subtropical gyres play a key role in the modulation of the carbon balance by seeding the inner ocean. To consider the role of this moving transitional area, we chose as analysis areas coastal strips of 8 degrees (in the meridional direction) in each system. However we used the full geographical areas to make our numerical computations. Note that the computation areas are larger than the analysis ones, considering the fact that particles may leave the area before reaching the fixed final distance $\delta_f$. In addition, several tests with different shapes and area selections (not shown) exhibit similar results.

## 3 Satellite data

A five year long time series from June 2000 to June 2005 of ocean colour data is used. Phytoplankton pigment concentrations (chlorophyll-$a$) are obtained from monthly SeaWiFS (Sea viewing Wide Field-of-view Sensor) products[1], generated by the NASA Goddard Earth Science (GES)/Distributed Active Archive Center (DAAC). The bins correspond to grid cells on a global grid, with approximately 9 by 9 km.

The weekly global $1/4°$ resolution product of surface currents developed by Sudre and Morrow (2008) has been used. The surface currents are calculated from a combination of wind-driven Ekman currents, at 15 m depth, derived from Quikscat wind estimates, and geostrophic currents computed from time variable Sea Surface Heights. These SSH were calculated from mapped altimetric sea level anomalies combined with a mean dynamic topography from Rio et al.

[1] We used the level 3 binned data with reprocessing 5.1. See http://oceancolor.gsfc nasa.gov for further details.

(2004). These weekly velocity data, which are then interpolated linearly to obtain a daily resolution with a 0.025° intergrid spacing, depend on the quality of their sources as the SSH fields and the scatterometer precision. However, they were validated with different types of *in situ* data such as Lagrangian buoys, ADCP and current meter float data. In our four areas, zonal and meridional components show respectively an average correlation coefficient ($R^2$) with for e.g. Lagrangian buoy data of 0.64 and 0.57.

We analyse satellite data which are two-dimensional fields. We are also interested in the third dimension and the influence of vertical movements in upwelling, which are known to be relatively intense. To perform this, we propose to compute the Ekman transport and the divergence of the velocity field from the available data. The Ekman transport was calculated using $U_E = \frac{T_y}{f\rho}$, where $T_y$ is the meridional wind stress (obtained from the Quikscat scatterometer weekly wind estimates), $\rho$ is the density of seawater and $f$ is the Coriolis parameter. We also look at the vertical dimension by quantifying the horizontal divergence of the surface velocity field, using the incompressibility assumption: $\Delta(x, y, t) \equiv \partial_z V_z = -(\partial_x V_x + \partial_y V_y)$. This calculation gives an estimate of the mean vertical velocities over the whole period. Negative (positive) values of $\Delta$ indicate upwelling (downwelling) areas because they signal surface spatial points where fluid parcels diverge (converge).

## 4 Results

### 4.1 Comparative study of the mixing activity

#### 4.1.1 Spatial distribution of the mixing properties from FSLEs

In Fig. 1 we draw the time average (covering the period June 2000–June 2005) of the FSLEs computed for the four EBUS. For all areas, two different subsystems, according to their mean mixing activity, can be defined. The zonal limits are as follows: 30° N for the Canary (CUS) and the California upwelling system (CalUS), 27° S for the Benguela (BUS) and 25° S for the Humboldt (HUS). Comparing these four upwelling zones, a distinction appears in both upwellings of the Southern Hemisphere where the limit between subsystems is clearly marked while in areas of the Northern Hemisphere it is not so sharp. Note also that the imaginary division lines are usually associated with particular topographic or hydrographic features: for the CUS, the line passes north of the Canary archipelago, the offshore HUS limit coincides with the Nazca Ridge and the BUS limit matches the latitude of the intense Lüderitz upwelling cell. The dark blue areas (FSLE value below 0.005 day$^{-1}$) located close to the coast indicate some retention zone since the mixing time is very large or infinite (the computation of FSLE gives zero when particles move toward the coast).

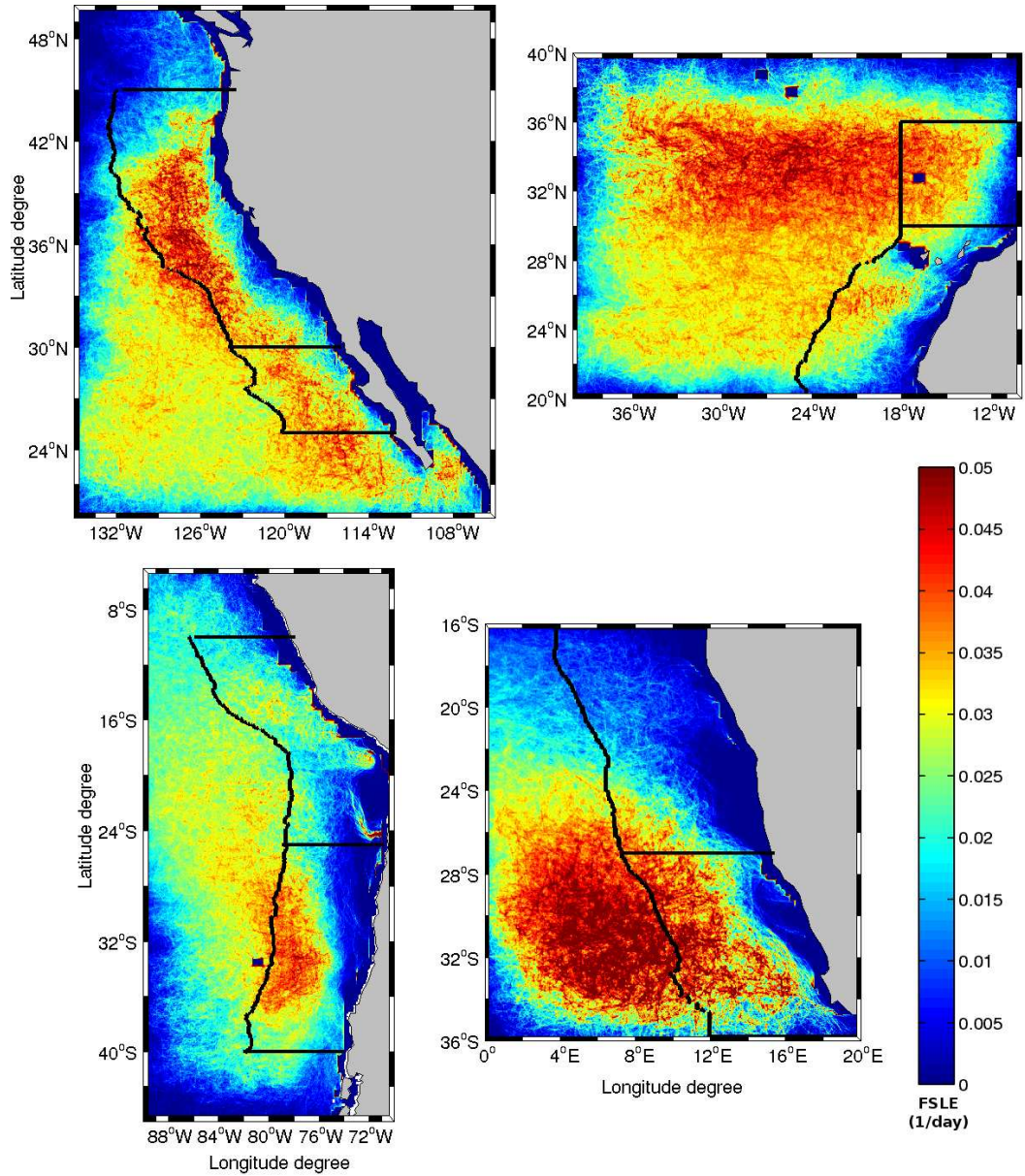

**Fig. 1.** Time average over the period June 2000–June 2005 of the FSLEs for the CalUS (upper left), the CUS (upper right), the HUS (lower left), and the BUS (lower right). Black lines indicate the analysis area as 8 degrees coastally oriented strips and the corresponding subdivisions.

To further quantify the variations in the stirring we examine the probability density functions (PDF) of FSLEs. These distributions are calculated for the FSLEs' time average normalized by the mean values from all grid points within each area (Fig. 2a). For all regions except the BUS (red line), the PDFs have a similar shape: their distributions are broad and slightly asymmetric, with a peak at low mixing activity and a quite long tail of high mixing. However the width and peak values vary depending on the considered system. The PDF of the BUS exhibits a particular asymmetric shape: we can ob-serve one high peak in the low FSLEs values (around half of the spatial mean value, corresponding to ∼0.008 day$^{-1}$) and a bump standing in moderate values of FSLEs (between 2 to 3 times the mean value, corresponding to 0.03–0.04 day$^{-1}$). Considering the very distinct PDFs of FSLEs between both BUS subsystems (as compared to the HUS) we can associate the high peak of low FSLEs to the northern subsystem, whereas the moderate FSLEs' bump constitutes a signature of the intense mixing in the southern subsystem constantly fed by numerous and powerful Agulhas rings. Note also that the CalUS exhibits a thinner and higher peak as compared to the others, indicating that the mean mixing is moderate and quite homogeneous over the entire analysis area (high occurrence of values close to 1, meaning many values are found around the spatial mean). Waugh and Abraham (2008) showed that the PDFs of FTLEs (for Finite-Time Lyapunov Exponents) have a near-universal distribution in the global open surface ocean since they are reasonably well fit by Weibull distributions following: $P(\lambda)=\frac{b}{a}\left(\frac{\lambda}{a}\right)^{b-1}\exp(\frac{-\lambda^b}{a^b})$, with $a=\bar{\lambda}/0.9$ and $b=1.6-2.0$. We expected a similar behavior for FSLEs because of the close relation among these quantities. We confirmed that normalized PDFs computed over the upwelling areas are quite well fitted by a Weibull distribution with parameters close to those proposed by these authors, except for the BUS. In Fig. 2b, the normalized PDFs of FSLEs from the CUS and HUS are quite well modeled by a Weibull distribution with parameter $b=2.2$ whereas the PDFs' from CalUS (Fig. 2c) fits better a Weibull distribution with parameter $b=3.4$ related with the higher and thinner peak around the average. The particular shape of the PDF of normalized FSLEs over the BUS indicates again that mixing in this upwelling system is much more heterogeneous.

### 4.1.2 Temporal evolution of the mixing intensity along the period 2000–2005

A more detailed comparison between the different subsystems can be performed by calculating the time evolution of the spatial averages over the analysis area of each of the four upwellings (Fig. 3a) and each subsystem (Fig. 3b and c). First of all we can sort each area according to their global averaged mixing activity. The mixing in the CalUS appears to be the most vigorous one (spatial average over the whole period: $0.025\,\text{day}^{-1}$), followed by the CUS ($0.021\,\text{day}^{-1}$), and finally the HUS ($0.019\,\text{day}^{-1}$) and BUS ($0.017\,\text{day}^{-1}$) which presents the lowest mixing activity. A strong annual signal is observed in the time evolution of the mixing in the Humboldt, Canary and California upwelling systems. The five peaks of high mixing, corresponding to the five years of data, reflect the seasonal variability of the surface wind. In each hemispheric winter, the sea surface exhibits a more turbulent behaviour due to stronger winds. The last year of these time series reveals a somewhat different pattern of mixing, with a double peak for both upwellings of the Northern Hemisphere, suggesting that 2005 might be a particular year. In fact, this event has been already documented by Schwing et al. (2006) who studied the large-scale atmospheric forcing that contributed to these unusual physical oceanic conditions and the associated ecosystems responses. Note that both systems of the Northern Hemisphere oscillate in phase and are out of phase with the Southern Hemisphere systems. Periods of minimum turbulence values, for instance in the HUS, occur from March through May and coincide with the upwelling relaxation period, linked with the coastal wind regimes. A similar observation may be done for the BUS: the minimum mixing values during austral summer/autumn coincide with the upwelling relaxation period. Despite the fact that a high variability of the mixing is observed in all systems, the Benguela exhibits the strongest interannual variability of the mixing among all four EBUS. Note also an increasing trend of the mixing in the CalUS over these five years, confirmed by the computation of FSLE anomalies (not shown), suggesting a long-term change linked with global climate change (Bograd et al., 2009, and references therein). In all four regions, the difference of horizontal mixing activity is clear between tropical and temperate subsystems which showed the maximum of mixing (Fig. 3b and c). This observation can be explained by the intensification of large scale atmospheric forcing at mid-latitudes. When going away from the relatively calm equator, the intensity of the trade winds is increased in the gyre, associated to the presence of jet streams and increased pressure gradient. As already mentioned, this difference is more pronounced in areas of the Southern Hemisphere than of the Northern one. In Fig. 1, one can see a weak predominance of red colour in the temperate subsystem of the CalUS (north), suggesting it might be the most turbulent one. However the picture is more complex due to the particular temporal evolution of their mixing activity. Initially slightly less active than the southern one, the northern subsystem exhibits a positive tendency of increase, whereas the former is characterized by a flat long term pattern. As a consequence, the temperate subsystem becomes more turbulent than the tropical one at around year 2003. These different behaviors of the northern and southern CalUS subsystems were recently studied by Bograd et al. (2009) using newly developed upwelling index. Finally, the CalUS is quite particular as compared to the others since its horizontal mixing activity is more homogeneous: when averaging it over space and time in each subsystem, the FSLE means are very comparable, the southern one being slightly higher. Comparing these four upwelling zones, one can note that in the most turbulent temperate subsystems the values of the FSLEs are quite similar: within the range 0.018–0.04 day$^{-1}$, i.e., horizontal mixing times between 40 and 90 days. On the contrary, the least active tropical subsystems (excluding Southern California) are characterized by FSLE ranges from 0.003–0.025 day$^{-1}$ equivalent to horizontal mixing times from 65 to 530 days. Again on Fig. 3b and c, the mixing activity of the four subsystems from the Northern Hemisphere seems to vary in phase and shows a minimum during the boreal summer/autumn. On the other hand, in the BUS and HUS, the most turbulent temperate systems exhibit a visible annual cycle, with a minimum occurring during the austral summer/autumn, whereas the least active tropical ones show a high non linear variability and no obvious trend. Note that the northern BUS shows the smallest mixing activity of all areas.

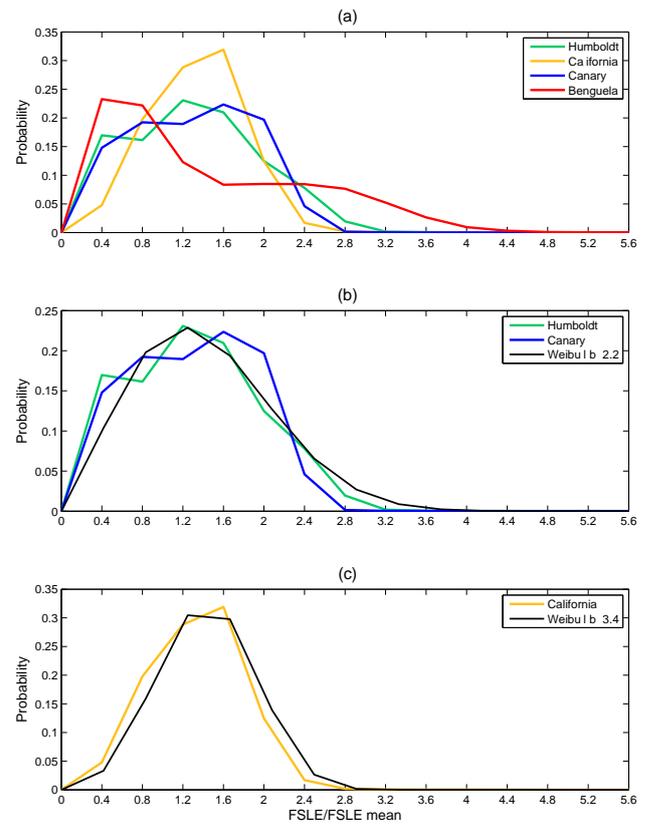

**Fig. 2. (a)** Normalized Probability Density Function calculated over the FLSEs time average of Fig. 1 for each EBUS (whole analysis area, i.e. 8 degrees coastal strip). Panels **(b)** and **(c)** same as in (a) for three of the PDFs fitting a Weibull distribution.

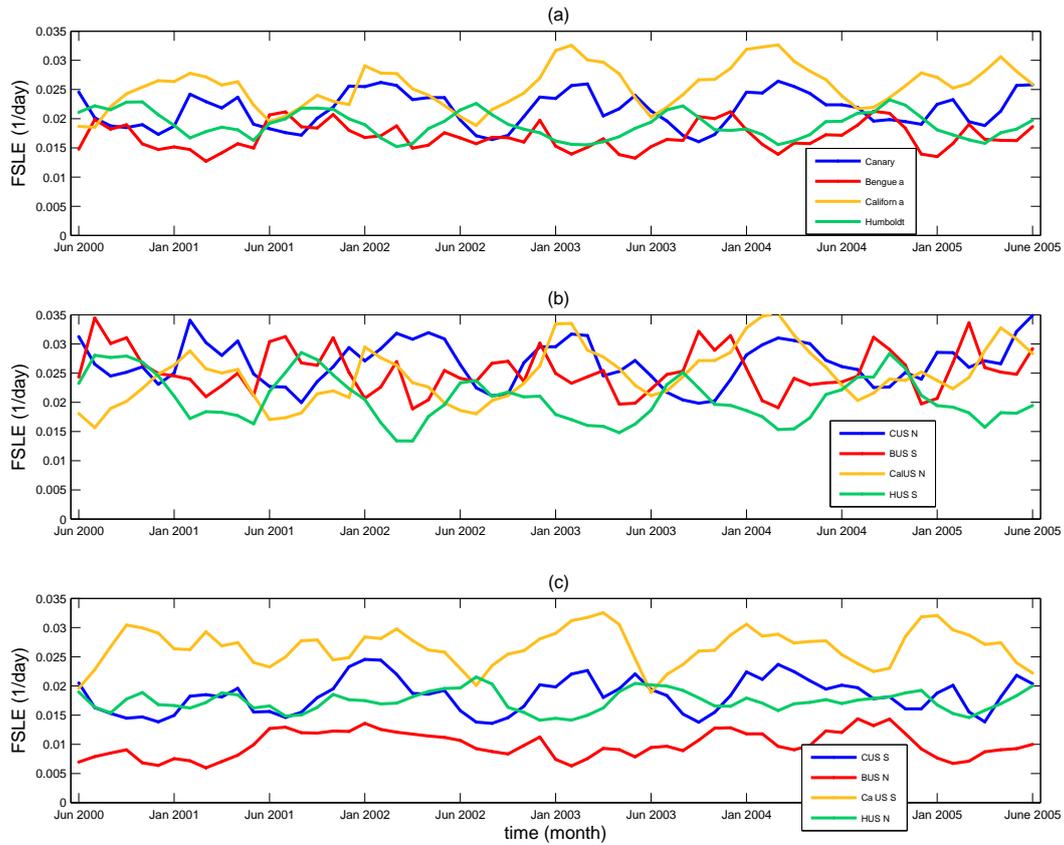

**Fig. 3.** **(a)** Spatial average versus time of the backward FSLEs. Spatial averages are computed over the analysis areas (8 degrees coastal strip): Canary (blue), Benguela (red), California (yellow) and Humboldt (green); **(b)** Same as in (a) but for the most temperate subsystems; **(c)** Same as in (a) but for the tropical subsystems.

The high spatio-temporal variability of the surface mixing revealed from FSLEs may strongly modulate the biological components of these complex and dynamic ecosystems. Next we proceed to investigate the correlation between horizontal mixing with the biological activity in our regions of interest.

### 4.2 Relationship with the biological activity

Now we study the relationship between the FSLEs and surface chlorophyll concentration estimated from space. First we performed Hovmöller plots of the surface chlorophyll distribution in the four upwellings by averaging the chlorophyll concentration along lines of constant latitude within the analysis areas for our five years of study (not shown). In each upwelling system, a clear distinction appears between two different zones, a southern one and a northern one, characterized by a very distinct degree of chlorophyll richness. In fact, the limits of the subsystems observed in the chlorophyll concentration Hovmöller plots coincide with the previous latitutinal limits deduced from FSLEs (around 30° N for CUS and CalUS, 27° S for BUS and 25° S for HUS). We also noticed that the poorest subsystem in chlorophyll matches the most turbulent one and vice-versa; this remark stands for the four EBUS. The spatial averaged chlorophyll over each analysis area (8° degrees coastal strips) reveals that the BUS admits the highest chlorophyll-*a* content (0.78 mg/m$^3$), followed by the HUS (0.43 mg/m$^3$), CUS (0.42 mg/m$^3$) and CalUS (0.36 mg/m$^3$). This ranking is just the opposite as the one based on the mixing activity of the surface ocean.

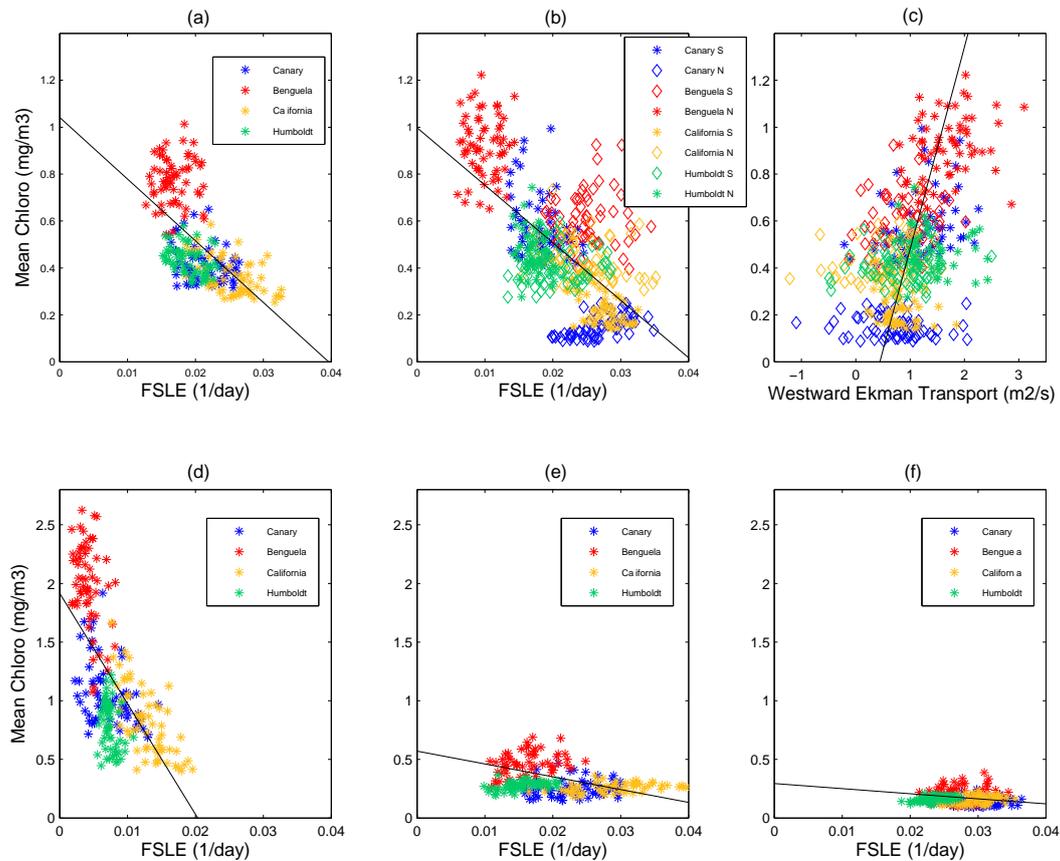

**Fig. 4.** Chlorophyll-*a* versus backward FSLEs, both averaged over the analysis areas (8 degrees coastal strips) for: **(a)** Whole analysis areas ($R^2$=0.38); **(b)** Same as in (a) but per subsystem ($R^2$=0.43); **(c)** Same as in (b) but for chlorophyll-*a* versus westward Ekman transport per subsystem ($R^2$=0.21; for visual improvement, the regression line has been obtained with the opposite order, x-axis versus y-axis); **(d, e, f)** Same as in (a) but for three successive strips oriented along the coast, **(d)** 2° from the coast, **(e)** within the 2° to 5° coastally oriented strip and **(f)** within the 5° to 9° coastally oriented strip.

If one plots spatial averages of FSLE versus spatial averages of chlorophyll concentration, over the entire analysis area and over each subsystem (Fig. 4a and b), for each month from June 2000 to June 2005, a negative correlation between FSLEs and chlorophyll concentration emerges. For all four areas, the subsystems with the highest mixing activity are the poorest in chlorophyll. Note on Fig. 4a that each upwelling system is characterized by a clear clustering, with an exception of the CalUS which presents a more widespread distribution. The same observation is valid for each subsystem on Fig. 4b. Although these EBUS present common features, they seem also to have particular functioning revealed by these clusters. When considering only the most turbulent subsystems of each EBUS, the correlation is rather clear ($R^2$=0.75, not shown) suggesting that the higher the mixing activity is, the stronger the inverse relationship is between surface chlorophyll and mixing. The inhibiting effect of stirring revealed by the negative correlation seems to occur above a certain value – a turbulence threshold – and might act differently depending on the system. Following similar calculations of spatial means over strips oriented along the coast (see Fig. 4, three lower panels d, e, and f), we observe a high negative correlation at the coast, decreasing when shifting to offshore strips, and even becoming flat when approaching the oligotrophic gyre further offshore. This finding obtained from an analysis over the four EBUS seems to indicate a variable response of the biology to physical stirring, valid in such diverse areas widespread over the world ocean.

Upwelling areas are definitely affected by water vertical movements and velocities, through uplift of rich nutrients water and 3-D turbulence, which are not captured by our previous analysis. We will also examine the influence of Ekman transport which creates pumping of nutrients and carries them from the deep layer to the coastal surface waters where light is not limiting. Vertical velocities and Ekman transport, which can both play a very relevant role in the chlorophyll signature detected from space, will be considered in the following.

We evaluated the horizontal divergence, $\Delta$, of the surface velocity field and averaged it over the period June 2000–June 2005 at each point of the CUS, BUS, CalUS and HUS (not shown). The negative values of the $\Delta$ field in the coastal areas indicates the presence of upwelling events. We noticed that in the coastal zones of the BUS, the well known upwelling cells Cape Frio, Walvis Bay and Lüderitz in the northern subsystem appear clearly, being more intense than the southern cells, in agreement with Monteiro (2009) estimates of the northern subsystem accounting for 80%, on average, of the total upwelled flux over the whole BUS. The intense upwelling cells spread along the Peru/Chile coast are also visible, whereas the area above 15° S is mainly characterized by negative velocities which represent the equatorial upwelling. When averaging the temporal mean of the $\Delta$ field over the analysis area of each subsystem, representing a measure of the mean vertical velocities averaged over space and time, we confirm that the less (most) horizontally stirred system is associated with negative (positive) mean vertical velocities indicating predominance of upwelling phenomena (downwelling, respectively). This stands for the BUS, HUS and CUS since their less turbulent systems are respectively characterized by $\overline{\Delta}_{nBUS}=-0.0036\,\text{day}^{-1}$, $\overline{\Delta}_{nHUS}=-0.002\,\text{day}^{-1}$ and $\overline{\Delta}_{sCUS}=-0.0016\,\text{day}^{-1}$ whereas their most turbulent exhibit positive means ($\overline{\Delta}_{sBUS}=0.0012$, $\overline{\Delta}_{sHUS}=2.2\times10^{-4}\,\text{day}^{-1}$ and $\overline{\Delta}_{nCUS}=8.7\times10^{-4}\,\text{day}^{-1}$). For the CalUS, the distinction of two different subregions is not so clear, as compared to the others, confirmed by the very close negative average $\overline{\Delta}_{nCalUS}=-8\times10^{-4}\,\text{day}^{-1}$ and $\overline{\Delta}_{sCalUS}=-1\times10^{-3}\,\text{day}^{-1}$. It seems that areas dominated with upward processes are restricted to the very coastal areas whereas the offshore waters are dominated by downward ones. The global averages of $\Delta$ over the whole domain reveal negative values (upwelling) and give the following ranking: the most intense upward velocities are found in the BUS, followed by the CalUS, then comes the HUS and finally the CUS.

To complete the analysis, we have calculated the Ekman transport $U_E$ along the E-W direction. Not surprisingly, its spatial distribution (not shown) is particularly linked to the spatial distribution of chlorophyll: high chlorophyll contents are often associated with intense Ekman transport, indicating high upwelling intensity. Indeed the northern regions of the BUS and HUS, the richest in chlorophyll and presenting the minimum mixing activity, are characterized by the highest offshore transport. In the CUS, both sub-areas have high values for the offshore transport very close to the coast, with similar values in the southern and northern subregions. Further from the coast, the highest westward transport in the southern CUS area coincides again with the highest chlorophyll content. The same analysis may be done for the CalUS with a less marked difference in the offshore southern subsystem. Figure 4c represents the spatially averaged westward Ekman transport $U_E$ versus spatial averages of chlorophyll concentration, over each subsystem from June 2000 to June 2005 (one point per month). Negative values indicate an Ekman transport to the east, whereas positive ones indicate an offshore Ekman transport to the west. A positive correlation appears confirming the effect of Ekman-transport induced upwelling on biological productivity. This finding is not surprising and compatible with previous results (Thomas et al., 2004) since horizontal currents are strongly related to the vertical circulation. A global average of Ekman transport over space (analysis areas) and time reveals similar ranking deduced from the chlorophyll content, except a shift between CUS and HUS. The BUS has the highest ($-1.33\,\text{m}^2/\text{s}$), then come the CUS ($-1.07\,\text{m}^2/\text{s}$) and HUS ($-1.01\,\text{m}^2/\text{s}$) and finally the CalUS ($-0.7\,\text{m}^2/\text{s}$).

## 5 General discussion

Divisions of each area into two subsystems, based on different levels of temporal averaged horizontal stirring rates, are quite consistent with limits deducted from different criteria in other studies (Carr and Kearns, 2003; Mackas et al., 2006; Monteiro, 2009). We also proposed a ranking of horizontal mixing that gives new insights as compared to the classification made by Patti et al. (2008) based simply on a wind-mixing index. Systems from the same hemisphere seem to exhibit a similar behaviour with a dominant annual cycle when studying the temporal evolution of their spatially averaged horizontal mixing activity. The study from PDFs confirmed the statistical structure of these Lagrangian diagnostics already documented by Waugh and Abraham (2008). They showed that PDFs of FTLEs (Finite-Time Lyapunov Exponents, comparable to FSLEs) computed over the global surface ocean may be reasonably modelled by two kinds of distributions: in weak strain regions they are well fitted by Rayleigh distributions whereas for large-strain regions PDFs are better fitted by a general Weibull distribution. Since most of the regions under study reveal a quite nice fitting to a Weibull distribution, with slightly different parameters values as indicated in Waugh and Abraham (2008), they can be considered as large-strain regions. The PDF of normalized FSLEs over the BUS shows a particular distribution, indicating that the mixing activity over this system is quite unique. The ranking in terms of chlorophyll content is the same than the one proposed by Cushing (1969), linking chlorophyll content and higher trophic level production. Carr (2001) and Carr and Kearns (2003) compared the EBUS depending on their primary productivity estimated from remote sensing and found also the same ranking, except for a switch between the HUS and CUS. The temporal variations of the chlorophyll stocks and their coupling with Ekman transport was studied in details by Thomas et al. (2004) over the four EBUS and more precisely over the CUS by Lathuilière et al. (2008). We globally confirmed that chlorophyll stocks are positively correlated with westward Ekman transport intensity.

When investigating the link of FLSEs with biological data, the scatterplots reveal a negative correlation between horizontal mixing activity and chlorophyll concentration in upwelling areas. This negative effect is in line with Lachkar et al. (2009) who showed that strong eddy activity acts as an inhibiting factor for the biological productivity in coastal upwelling systems. They confirmed that the CUS and CalUS appear to be the most contrasting systems of the 4 EBUS, in terms of biological productivity and mixing activity as well. Patti et al. (2008) also mentioned a negative correlation between turbulence, calculated as the cube of wind speed, and logarithm of the chlorophyll-$a$ concentration for the BUS; however this finding did not hold for the other areas showing a positive relationship. We note that theoretical studies in idealized settings, in which nutrients reach plankton only by lateral stirring, display also negative correlation between mixing and biomass (although mixing and productivity may be positively correlated) (Tél et al., 2005; Birch et al., 2007; McKiver and Neufeld, 2009). In the following we propose some mechanisms to explain this inverse relationship, as compared to the open ocean and other low nutrient environments, where several studies showed that eddies and turbulent mesoscale features tend to rather enhance biological productivity (McGillicuddy et al., 1998; Oschlies and Garçon, 1998).

In our case, we focussed on very productive areas where the high biological productivity is maintained by a large nutrient supply from deep waters driven by Ekman pumping. Horizontal turbulent mixing of nutrients in surface waters, which was significant in the oligotrophic areas, is now a second order effect as compared to the vertical mechanisms (nutrient Ekman pumping) in the most productive subsystems. McKiver and Neufeld (2009), lay the emphasis on a ratio between the biological ecosystem timescale (inverse of the growth rate) and the flow timescales. When increasing the ratio, corresponding to an increase of turbulence, they indicate a negative effect on the phytoplankton mean concentration as it is the case in our study. The localized pulses of nutrients are rapidly being dispersed by intense mixing before being used efficiently by the phytoplankton to grow. Similar processes were documented in a theoretical modeling study from Pasquero et al. (2005). When associating an upwelling of nutrient with coherent vortices, they find a lower primary productivity than without vortices. They explained this observation by the trapping properties of eddies and the limited water exchange between the vortex cores and the surrounding waters. Eddies are able to trap and export offshore rich coastal waters which are not being used efficiently by the phytoplankton, resulting in a lateral loss of nutrients of the coastal upwelling. We also observed that areas characterized by high FSLEs are correlated with intense vertical movements (downwellings as well as upwellings), whereas the areas with low FSLEs are mainly dominated by upward vertical velocities (upwellings). Lehahn et al. (2007) recently showed that vertical motions associated with eddy are more precisely located close to the lines of high FSLEs. Regions of high FSLE averages indicate a high occurrence of intense eddies which modify the three dimensional mean flow. The nutrient Ekman pumping, dominant process in upwelling areas, is weakened and the fuelling of nutrients toward the surface decreased. A significant stirring revealed by high FSLEs may decrease the Ekman transport induced upwelling leading to weaker surface chlorophyll stocks.

In the scope of previous works concentrating in the open ocean, and considering Fig. 4 (3 lower panels), we suggest the possibility of a variable response of the phytoplankton to the mesoscale oceanic turbulence. This changing behavior, represented by the high negative correlation at the coast decreasing when moving offshore, may be explained when considering the different dominant processes in the areas of interest. In very coastal areas, intense biological productivity is supported by the intensity of Ekman induced upwelling of nutrients. However, a high turbulence caused by eddies may induce an offshore lateral loss of nutrients and may decrease the vertical fuelling of surface waters from nutrients Ekman pumping, thus leading to a negative effect on biological production. Then, when moving offshore in the transitional area between the very coastal upwelling and the oligotrophic gyre, the moderate production regime relies on the offshore export of coastal rich waters. In this case, the turbulent mixing of nutrients may have a minor influence on moderate productivity, from the compensation of weak positive and negative effects. Then, in the open ocean where the biological productivity is weak and limited by very low nutrient concentrations, the resultant effect of horizontal mixing on phytoplankton growth becomes positive. The phytoplankton development is being promoted by eddies which induce vertical velocities and an upward flux of nutrients toward the very depleted surface waters (McGillicuddy et al., 1998; Oschlies and Garçon, 1998).

Other factors are of course influencing at different levels the biological productivity in the ocean, more particularly in these highly variable areas. Several studies tried to iden-tify the main factors among the four EBUS. Carr and Kearns (2003) distinguished different types of factors and discussed which ones control primary productivity. Oxygen concentrations and displacement of the thermocline symbolized the large-scale upwelling intensity; the local forcings were represented by quantities such as Photosynthetically Available Radiation, offshore transport and SST gradient, but the authors omitted the turbulence. They showed that large-scale circulation patterns are responsible for the main differences between EBUS. Then, the local forcings, and their combination with large-scale factors, explain the weaker variations. If we would consider only the nutrient concentration and Ekman pumping intensity (from their study) on one hand, and the turbulence from FSLEs (from our study) on the other hand, we can easily explain differences among EBUS without taking into account all other factors. Here we argue that adding our turbulence data from FSLEs to nutrients concentration and Ekman transport intensity allow us to simply obtain similar results, suggesting the fact that the turbulence effect is important to be considered. Patti et al. (2008) studied the factors driving the chlorophyll content and they found that nutrient local concentrations, mainly governed by local upwelling intensity (Ekman pumping), explain the main differences between very productive areas (HUS and BUS) as compared to the other two (CalUS and CUS). These processes act as first order factors whereas the continental shelf width appears to be the key secondary order factor explaining the difference between HUS and BUS, also mentioned by Carr and Kearns (2003). The mixing from FSLEs can also explain the main differences observed since the CalUS and CUS admit the highest surface mixing activity. Moreover, the highest chlorophyll content observed in the northern Benguela coincides again with the minimum of mixing measured by FSLEs. The BUS appears to be the most productive system since the Ekman pumping over a large width shelf is maximum and associated with the lowest mixing. Patti et al. (2008) also discussed other factors such as light limitation, solar cycle, presence of retention areas, etc., concluding that they should act at different levels, in different areas. It is also well known that micronutrient availability and alternative biogeochemical processes such as $N_2$ fixation or denitrification may also play a role in nature. However, the variability among all these factors over all areas is too large to identify trivial patterns. Consequently we did not consider these controls as primary factors in our analysis.

## 6 Conclusions

The distribution of FLSEs computed from multi-sensor velocity fields over a 5 year period allowed us to make a comparative study of the mixing activity in the four eastern boundary upwelling systems. Each area was divided into two subsystems showing different levels of temporal averaged horizontal stirring rates. When studying the temporal evolution of their spatially averaged mixing, we proposed a ranking in terms of horizontal mixing intensity for all four EBUS. We also found that the more vigorous mixing occurs in subsystems further away the equator explained by the intensification of large scale atmospheric forcing at higher latitudes. Systems from the same hemisphere seem to exhibit a similar behavior with a dominant annual cycle. The PDF computations of FSLEs reveal the statistical structure of these Lagrangian diagnostics. When investigating the link of FSLEs with biological data, the subdivisions detected from FSLEs' maps appeared to be also visible on chlorophyll concentration Hovmöllers suggesting that these two quantities are linked. The scatterplots revealed a negative correlation between horizontal mixing activity and chlorophyll concentration in upwelling areas. We then confirmed that chlorophyll stocks are positively correlated with westward Ekman transport intensity over the four EBUS. It thus seemed that the horizontal turbulent mixing of nutrient is a second order effect as compared to the vertical mechanisms. After estimating the mean vertical velocities from incompressibility assumption, we proposed another explanation: the regions of high FSLEs are characterized by occurence of intense eddies and their verticals velocities associated. This will modify the whole 3-D flow and lead to a global decrease of the nutrient Ekman pumping (supported by low Ekman transport). We finally suggest the possibility of a variable response of the phytoplankton to the sub/mesoscale oceanic turbulence depending on the distance to the coast. This changing behavior is represented by the high negative correlation at the coast decreasing when moving offshore. It may be explained when considering the different areas and their associated dominant bio-physical processes. We then discuss the effect of others factors not considered here, and compare our approach to all previous comparative works.

Further work should investigate the robustness of the relationship found in our four systems when examining FSLEs versus biological stocks. Still much needs to be done to fully understand how plankton distributions are controlled by the interplay between their turbulent medium and the non-linear processes of their ecology. Coupled modelling approaches appear to be the only way to consider all these factors simultaneously. Besides a better understanding of the interactions between biological and physical processes, these coupled modelling studies will allow us to investigate and determine the respective effect of abiotic and biotic factors.

Although dealing with other scales of study may lead to different conclusions, chaotic stirring and turbulence in the ocean play a very important role by influencing biological processes at any scale. The negative effect of horizontal stirring on biological productivity in upwelling areas shown here needs to be considered when trying to estimate the carbon-pump efficiency on a global scale since upwelling areas shelter more than 20% of the global biological productivity (Ryther, 1969; Cushing, 1969; Chavez and Toggweiler, 1995). The global estimation of $CO_2$ fluxes at the ocean-atmosphere interface will gain in accuracy when considering this effect through, for instance, a spatial parameterization of turbulence.

*Acknowledgements.* V. R. and C. L. were awarded a EUR-OCEANS network of Excellence short visit grants. V. R. is supported by a PhD grant from Direction Générale de l'Armement. V. G. and J. S. acknowledge funding support from CNES and C. L. and E. H.-G. from PIF project OCEANTECH of the Spanish CSIC and FISICOS (FIS2007-60327) of MEC and FEDER. We also thank A.M. Tarquis and the two anonymous referees for their constructive comments.

Edited by: A. Turiel
Reviewed by: A. M. Tarquis and two other anonymous referees

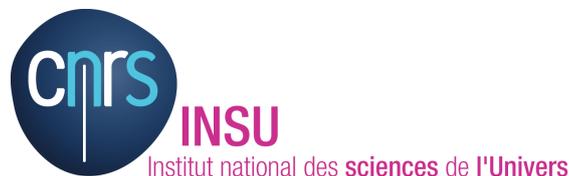

The publication of this article is financed by CNRS-INSU.

568 V. Rossi et al.: Surface mixing and biological activity in the four Eastern Boundary Upwelling Systemsmary production in a model of the North Atlantic Ocean, Nature, 394, 266–268, 1998.

Owen, R. W.: Fronts and Eddies in the sea: mechanisms, interactions and biological Effects, in: Fronts and Eddies in the Sea, p 197–233, edited by: Owen, R.W., Academic Press, London, 1981.

Patti, B., Guisande, C., Vergara, A.R., Riveiro, I., Barreiro, A., Bonanno, A., Buscaino, A., Basilone, G., and Mazzola, S.: Factors responsible for the differences in satellite-based chlorophyll *a* concentration between the major upwelling areas, Est. Coast. Shelf Sc., 76, 775–786, 2008.

Pasquero, C., Bracco, A., and Provenzale, A.: Impact of the spatiotemporal variability of the nutrient flux on primary productivity in the ocean, J. Geophys. Res., 110, C07005, doi:10.1029/2004JC002738, 2005.

Pauly, D. and Christensen, V.: Primary production required to sustain global fisheries, Nature, 374, 255–257, 1995.

Rio, M.-H. and Hernández F.: A mean dynamic topography computed over the world ocean from altimetry, in-situ measurements, and a geoid model, J. Geophys. Res., 109, C12032, doi:10.1029/2003JC002226, 2004.

Rossi, V., López, C., Sudre, J., Hernández-García, E., and Garçon, V.: Comparative study of mixing and biological activity of the Benguela and Canary upwelling systems, Geophys. Res. Lett., 35, L11602, doi:10.1029/2008GL033610, 2008.

Ryther, J. H.: Photosynthesis and fish production in the sea, Science, 166, 72–76, 1969.

Schwing, F. B., Bond, N. A., Bograd, S. J., Mitchell, T., Alexander, M. A., and Mantua, N.: Delayed coastal upwelling along the U.S. West Coast in 2005: A historical perspective, Geophys. Res. Lett., 33, L22S01, doi:10.1029/2006GL026911, 2006.

Shadden, S. C., Lekien, F., and Marsden, J. E.: Definition and properties of Lagrangian coherent structures from finite-time Lyapunov exponents in two-dimensional aperiodic flows, Phys. D, 212, 3–4, 271–304, 2005.

Sudre, J. and Morrow, R.: Global surface currents: a high resolution product for investigating ocean dynamics, Ocean Dyn., 58(2), 101–118, 2008.

Tél, T., de Moura, A., Grebogi, C., and Karolyi, G.: Chemical and biological activity in open flows: A dynamical system approach, Phys. Rep., 413, 91–196, 2005.

Thomas, C. A., Strub, T. P., Carr, M. E., and Weatherbee, R.: Comparisons of chlorophyll variability between the four major global eastern boundary currents, Int. J. Rem. Sens., 25, 7, 1443–1447, 2004.

Waugh, D. W. and Abraham, E. R.: Stirring in the global surface ocean, Geophys. Res. Lett., 35, L20605, doi:10.1029/2008GL035526, 2008.
Nonlin. Processes Geophys., 16, 557–568, 2009   www.nonlin-processes-geophys.net/16/557/2009/